\documentclass{aa}
\usepackage{psfig}
\def\micron{\hbox{$\mu$m}} 
\begin{document}

   \thesaurus{11     
              (11.11.1;  
               11.14.1;  
               11.09.1 M\,83; 
               11.19.2;  
               11.19.3;  
               11.19.5;  
               11.19.6;)}  

   \title{Stellar dynamics observations of a double nucleus in M\,83}

   \author{N. Thatte
           \inst{1}
           M. Tecza
           \inst{1}
           \and
           R. Genzel
           \inst{1}
           }

   \offprints{N. Thatte}

   \institute{Max Planck Institute f\"ur extraterrestrische Physik, 
              Giessenbachstra\ss e, D-85748 Garching, Germany\\
              e-mail: thatte@mpe.mpg.de
             }

   \date{Received August 16, 2000}

   \maketitle 

   \begin{abstract} 

   We report on the discovery of a double nucleus in M\,83, based on
   measurements of the line of sight velocity distribution of stars
   observed at near infrared wavelengths with the VLT ISAAC
   spectrograph.  We observe two peaks separated by 2\farcs 7 in the
   velocity dispersion profile of light from late-type stars measured
   along a slit 0\farcs 6 wide, centered on the peak of K band
   emission and with P.A. 51.7\degr .  The first peak coincides with
   the peak of the K band light distribution, widely assumed to be the
   galaxy nucleus.  The second peak, of almost equal strength, almost
   coincides with the center of symmetry of the outer isophotes of the
   galaxy.  The secondary peak location has little K band emission,
   and appears to be significantly extincted, even at near infrared
   wavelengths.  It also lies along a mid-infrared bar, previously
   identified by Gallais et al. (1991) and shows strong hydrogen
   recombination emission at 1.875\micron .  
   If we interpret the observed stellar velocity dispersion as coming
   from a virialized system, the two nuclei would each contain an
   enclosed mass of 13.2 $\times$ 10$^6$M$_{\sun}$ within a radius of 5.4
   pc.  These could either be massive star clusters, or
   supermassive dark objects. 

   \end{abstract}

%

\section{Introduction}

M\,83 is a very nearby grand design barred spiral (Hubble type
SAB(s)c, distance 3.7 Mpc, de Vaucouleurs et al. 1991) showing
vigorous star forming activity in its nuclear region.  It has been the
object of numerous studies at all wavelengths, ranging from the X-ray
(\cite{immler}), visible (\cite{sofue}; \cite{comte}), near-infrared
(Gallais et al. 1991), mid-infrared (Telesco, Dressel and Wolstencroft
1993; Rouan et al. 1996), to the radio (\cite{ishizuki}).
Significant dust extinction in the nuclear region (\cite{turner})
implies that the true morphology of the starburst is revealed only at
infrared and longer wavelengths.

Although the morphology of the nuclear star forming activity has been
studied in detail, very few kinematic or dynamical studies at arc
second resolution have been carried out.  Radio measurements exist,
but are hampered by the large beam size. Puxley, Doyon and Ward (1997)
have performed low resolution near infrared spectroscopy along a
single slit, characterizing the stellar population in a broad way.  As
part of a program to search for supermassive black holes in nearby
spiral galaxies, we have conducted long slit medium resolution
spectroscopy of the nuclear region of M\,83 using the ISAAC near
infrared spectrometer on the European Southern Observatory's (ESO)
Very Large Telescope VLT.  Our observations are suggestive of the
presence of a double nucleus.

\section{Observations and Data Reduction}

\subsection{NTT SOFI imaging}
As part of the near infrared imaging carried out for the supermassive
black hole search pilot program, we obtained K band images of the
central region of M83 using the SOFI near infrared camera on the ESO
NTT telescope.  M\,83 was observed on 14-15 February 2000 for a total
of 600 seconds with the K$_{\rm s}$ filter.  Individual exposure times
were 10 seconds long, grouped together in sets of 12 exposures.  Three
exposures on blank sky located 5\arcmin\ E of the galaxy nucleus were
interposed after every group of on-target exposures.  A jitter pattern
with a 20\arcsec\ throw in both R.A. and Dec was used to offset every
group of on-target exposures relative to the previous one.  The seeing
was 0\farcs 9 during the
observations.  We used the large field mode of SOFI, with a pixel
scale of 0\farcs 29 and a total field of view of $\sim$5\arcmin
$\times$ 5\arcmin .  The reference star GSPC S791-C was observed a few
minutes later for use as a photometric reference.

\begin{figure}
\begin{center}
\psrotatefirst
\psfig{figure=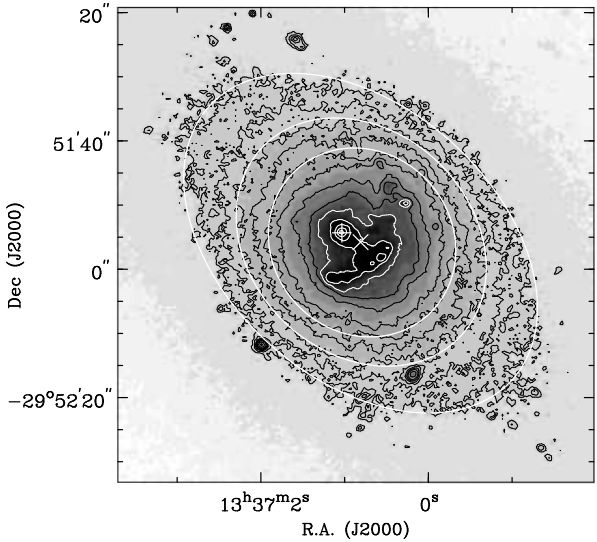,width=8.8cm,angle=-90}

\caption[]{K$_{\rm s}$ image of the central region of M\,83, obtained
with SOFI at the ESO NTT.  Grey scale and contour representations are
superposed on each other.  Contours range from the peak intensity to 5
mag below the peak in steps of 0.25 mag.  The plus indicates the
location of the photometric peak, while the cross indicates the center
of symmetry of the outer isophotes of the galaxy.  Fits to the outer
isophotes are indicated by white ellipses.}
\label{sofiimage}
\vspace{-1cm}
\end{center}
\end{figure}

Data reduction was carried out using the ECLIPSE\footnote{The ECLIPSE
package ({\em www.eso.org/eclipse}) 
consists of a set of stand alone routines provided by ESO for
SOFI and ISAAC data reduction} (\cite{devillard}) jitter routines.
The process consists of applying dark frame subtraction, flat field
division and bad pixel correction to every group of exposures.  The
source and sky exposures are then input to a dejittering routine which
performs sky subtraction, determines the offset between groups and
re-centers frames, and finally co-adds the frames to yield the reduced
image.  Figure \ref{sofiimage} shows the K$_{\rm s}$ image of the
central region of M\,83, as observed with SOFI. 

\subsection{VLT ISAAC Spectroscopy}

We obtained spectra of the nuclear region of M\,83 on March 20--21,
2000, using the near infrared spectrometer ISAAC at the VLT.  We used
two slit positions at position angle (P.A.) 51.7\degr\ (major axis)
and 141.7\degr\ (minor axis).  The major axis slit is located along
the primary stellar bar whose position angle is measured from the SOFI
images.  The slit width was set to 0\farcs 6 ($\sim$ 4 detector
pixels), corresponding to an instrumental resolution of (R $\equiv$
$\lambda/\Delta\lambda$) 4750.  The spectral coverage extended from
2.238\micron\ to 2.361\micron , with a scale of $\Delta\lambda =$
1.212\AA\ per pixel.  We observed for a total on-source integration
time of 4500 seconds per slit setting.  Individual exposures were 300
seconds long. We nodded 30\arcsec\ along the slit after each on-target
exposure.  Due to the large spatial extent of the galaxy which spanned
the entire length of the ISAAC slit (120\arcsec ), we nodded the
telescope to blank sky 5\arcmin\ E of the galaxy nucleus after every
three on-target exposures.  The seeing during the observations varied
between 0\farcs 7 and 1\farcs 1.  The star HD 118187 (F7/F8V) was
observed as a spectroscopic calibrator on both nights, interleaved
with the observations of M\,83.

In addition, we obtained spectra of several late type giant stars
(spectral types K1III to M5III) for use as spectral templates in the
data analysis.  Each star was observed using the same slit setting as
M\,83, and nodded 30\arcsec\ along the slit every 5 to 10 seconds.
Details of the template stellar spectra will be presented in a
subsequent paper.

We dark subtracted, flat-fielded and dead pixel corrected each
spectroscopic exposure, and then transformed it onto a linear
wavelength grid with dispersion axis precisely parallel to detector
rows.  Our sky subtraction was done using {\em smoothed sky} exposures
so as not to degrade the signal to noise ratio (SNR).  We smoothed
each sky spectrum over 100 spatial pixels while maintaining the small
scale spatial structure along the slit, to produce the smoothed sky.
The results yielded useful galaxy spectra over the entire slit length
with better SNR than using standard {\em nodding-on-slit} techniques.
Atmospheric transmission corrections were made by dividing the sky
subtracted, co-added galaxy spectra by spectra of the spectroscopic
reference star, thus eliminating the only significant telluric
absorption feature at 2.317\micron .  Finally, a second interactive
dead pixel correction was employed to tag any hot or transient pixels.
The continuum for each spatial pixel along the slit was normalized to
unity over the wavelength range 2.255 -- 2.300\micron , excluding the
Ca feature at 2.265\micron .

The template star spectra were reduced in a manner similar to the
galaxy spectra, with the exception that sky subtraction was performed
by subtracting two nodded exposures from each other, shifting the
positive maxima so that they overlay, and co-adding the data.  A
single spectrum was extracted for each template star using the {\em
apall} routine within the {\em twodspec} package of IRAF.

\subsection{Archival HST NICMOS images}

We supplemented our broad band ground based images and medium
resolution spectra with archival data from the HST NICMOS
camera. (P.I. M. Rieke, Proposal I.D. 7218).  In particular, we used
6 datasets obtained with the NIC2 camera, taken on 16 May 1998.
Table \ref{nicmosdata} lists the details of the archival data used.
The NIC2 camera has a pixel scale of 0\farcs 075, corresponding to a
total field of view of 19\farcs 2 $\times$ 19\farcs 2.  The
observations were made in mosaic mode using multiple readouts of the
detector for each pointing. 

\begin{table}
\begin{center}
\caption[]{Summary of NICMOS archival data}
\label{nicmosdata}
\[
\begin{tabular}{|c|c|c|c|c|}
\hline
\noalign{\smallskip}
Dataset name & Filter & T$_{\rm int}$ & Feature \\
\noalign{\smallskip}
\hline
\noalign{\smallskip}
N4BV100D0\_MOS & F160W & 48 s& H band \\
N4BV100K0\_MOS & F187N & 160 s& Pa $\alpha$ \\
N4BV100N0\_MOS & F190N & 160 s& Pa $\alpha$ continuum \\
N4BV100G0\_MOS & F212N & 576 s& H$_2$S(1) \\
N4BV100T0\_MOS & F215N & 576 s& H$_2$S(1) continuum \\
N4BV100X0\_MOS & F222M & 176 s& K band \\
\noalign{\smallskip}
\hline
\end{tabular}
\]
\end{center}
\end{table}

\begin{figure}
\begin{center}
\psrotatefirst
\psfig{figure=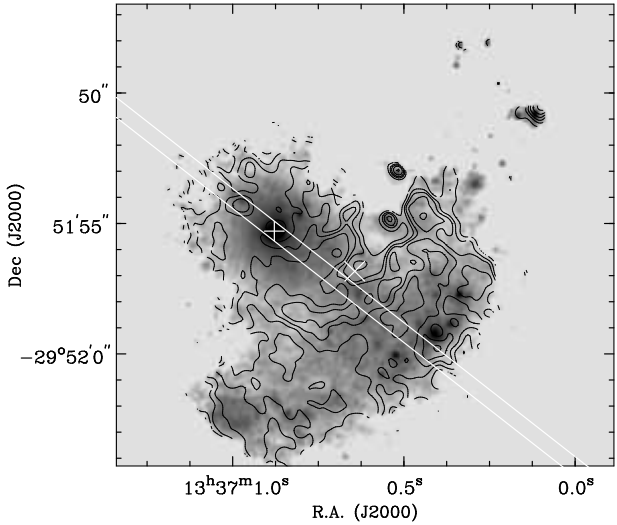,width=8.8cm,angle=-90}
\caption[]{NICMOS 2.22\micron\ image of the nuclear region of M\,83
(grey scale in magnitude units), with contours representing the H-K
colour overlaid.  Contours are at 0.2,0.3,0.4,0.5,0.6 and 0.8
magnitudes.  The locations of the major axis slit used for ISAAC
spectroscopy, are also indicated.  A cross indicates the location of
the center of symmetry of the outer isophotes of the galaxy.}
\label{nicmosmap}
\vspace{-1cm}
\end{center}
\end{figure}

We used pipeline calibrated data from the STSCI archive, processed by
both the {\em calnica} and {\em calnicb} pipelines.  These pipelines
together provide dead pixel and cosmic ray correction, flat fielding,
dark subtraction, photometric calibration and co-addition of individual
pointings of a mosaic into a single image.  We used the photometric
F$_\nu$ calibration, together with H and K zero points (\cite{cox}) to
establish a flux scale (in magnitudes) for the F160W and F222M images
and to yield a H-K color map.  Figure \ref{nicmosmap} shows the F222M
NICMOS image of the nuclear region of M\,83, with the H-K color
overlaid in contours.  Both the H and K images were smoothed prior to
subtraction with a two dimensional Gaussian with $\sigma =$ 0\farcs
150, so as to mitigate artifacts arising from the different point
spread functions at the two wavelengths.  The resulting H-K map still
shows a residual weak depression in the H-K color at the location of every
strong point source.

\section{Analysis and Results}

\begin{figure*}
\label{rawspectra}
\begin{center}
\vbox{\psrotatefirst
\psfig{figure=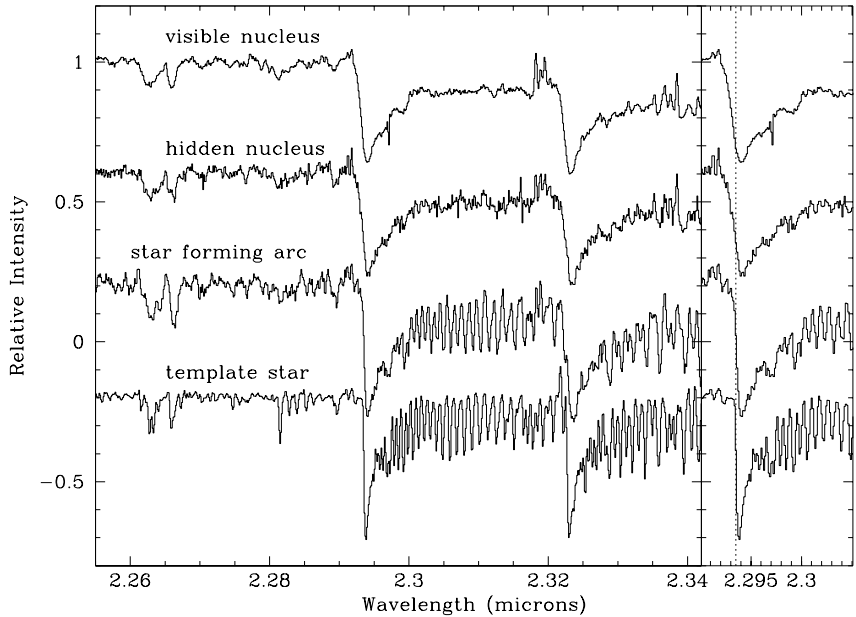,width=12cm,angle=-90}\vspace{-7cm}}
\hfill\parbox[b]{5.5cm}{
\caption[]{ISAAC spectra of the visible nucleus, hidden nucleus, star
forming arc and a template star of type K5III (top to bottom), each
within a 0\farcs 6 aperture.  The galaxy spectra have been shifted to
zero redshift.  Every spectrum has an amplitude offset of -0.4
relative to the one above it.  The resolved CO lines between 2.30 and
2.32\micron\ are very sensitive indicators of velocity dispersion.  An
expanded view of the region around the $^{12}$CO bandhead at
2.293\micron\ is shown on the right.  The velocity broadening toward
both nuclei is clearly seen.}}
\vspace{0.5cm}
\end{center}
\end{figure*}

\subsection{Isophote fits: the location of the photometric centroid} 
We used the ELLFIT routine with the GIPSY\footnote{The Groningen Image
Processing SYstem} (\cite{gipsy}) package to
perform elliptical isophote fits to the K band surface brightness
distribution observed with SOFI.  We fit points within a magnitude 
interval ranging from 5.00 to 3.75 of the peak intensity, with a width
of 0.25 magnitudes for each fit, and a step size of 0.25 magnitudes.
All fit parameters for the ellipses were unconstrained.
Figure \ref{sofiimage} shows some of the isophotal fits overlaid on a
contour and grey scale image of the central region of M\,83.  The
center of symmetry obtained from the fits lies 1\farcs 54 $\pm$
0\farcs 27 South and 3\farcs 05 $\pm$ 0\farcs 30 West of the
photometric peak.  
\cite{wolstencroft} also observed a similar offset (3\arcsec at
P.A. 255\degr ) based on lower resolution K band data.  
At larger spatial scales ($\sim$ 2\arcmin ), the bar is easily
identified in the K band image.  We measure the positional angle of
the bar from the SOFI image to be 51.7\degr $\pm$ 1\degr . 

\subsection{Recession velocity and velocity dispersion: discovery of a
hidden nucleus}
We used both Fourier correlation quotient and $\chi^2$ fitting techniques to
determine the recession velocity and the velocity dispersion as a
function of slit position.

The reduced ISAAC data provided a normalized spectrum for each spatial
pixel along the slit.  We binned the spectra along the spatial
direction with a binning width of four pixels (equal to the slit
width), yielding spectra with enhanced SNR spaced every 0\farcs 6
along the slit.   The total spatial extent of the major axis slit 
used for dynamical analysis was $\sim$ 13\arcsec , based on an SNR
cutoff of 20 per pixel in the continuum. Figure \ref{rawspectra} shows
some representative galaxy spectra around the region of the CO
0$\rightarrow$2 bandhead feature at 2.29\micron , as well as a
spectrum of a template star for comparison.  

We measured the redshift and velocity dispersion as a function of
spatial position along the slit by fitting a suitably broadened and
shifted template stellar spectrum to the galaxy spectrum.  The
normalized galaxy and template star spectra were resampled onto a grid
with equispaced velocity intervals (corresponding to logarithmic
intervals in wavelength).  The template star spectrum was convolved
with a Gaussian broadening function whose width and mean velocity were
free parameters.  A least squares fit, minimizing the normalized mean
square error, $\chi^2$, between the broadened template star spectrum
and the galaxy spectrum yielded the best fit values for the recession
velocity and velocity dispersion.  The range of the fit was optimized
to include all significant spectral features (CO bandheads and Ca
absorption lines) and a minimum of line-free continuum.  The lack of
continuum longward of the CO bandhead features can lead to incorrect
estimates of the continuum level.  Consequently, we also included a
quadratic polynomial term to account for a mismatch in the continuum
levels of the two spectra.

A thorough error analysis was carried out estimating both systematic
errors (due to template mismatch) and random errors resulting from
noise in the galaxy spectra.  Systematic errors were estimated using
template stars of varying spectral type (K3III to M2III) for the fit
and measuring the observed differences in fit velocities and
dispersions.  Random errors were estimated using a reduced $\chi^2$
technique.  Artificially broadened template star spectra with added
random noise (to appropriately mimic the observed SNR of the galaxy
spectra) were fitted as described above.  The change in fit parameters
required to increase the reduced $\chi^2$ by unity is a measure of the
error in each fit parameter.  The instrumental resolution corresponds
to a $\sigma$ of 27 km\,s$^{-1}$, and represents the limit of what can
be resolved by these observations.

\begin{figure}
\begin{center}
\psrotatefirst
\psfig{figure=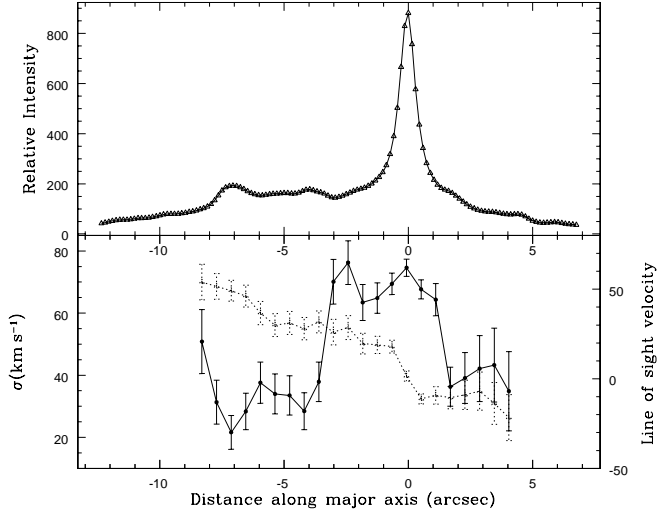,width=8.8cm,angle=-90}
\caption[]{Photometric and kinematic profiles along the major axis of
M 83.  The top plot shows the intensity for a 0\farcs 6 slit at
P.A. 51.7\degr\ as a function of offset from the location of the
visible nucleus. Positive offsets are toward the north-east.  The
bottom plot shows the recession velocity (dotted line and triangles)
and line of sight velocity dispersion (solid line and circles) as
determined by comparison with a stellar template over the wavelength
range 2.26 -- 2.34 \micron , using a $\chi^2$ fit.  Error bars
represent 3$\sigma$ uncertainties including both systematic and random
errors.  The secondary nucleus is evident as a second dynamical peak
which roughly matches the dip in the K band light profile.  The
visible nucleus is also associated with a sharp gradient in the
recession velocity.  }
\label{velprofile}
\vspace{-1cm}
\end{center}
\end{figure}

Figure \ref{velprofile} shows the results of the $\chi^2$ fitting for
the major axis of M\,83.  A velocity dispersion peak is coincident with
the location of the K band photometric peak, believed to be the
nucleus.  We also observe a second velocity dispersion peak, of equal
strength, offset 2\farcs 7 south-west from the nucleus (see figure
\ref{rawspectra} for a comparison of broadened CO bandhead profiles at
the two peaks). The photometric peak is also associated with a sharp
gradient in recession velocity, while no such jump is observed at the
location of the secondary peak.  The off-nuclear velocity dispersion
peak could possibly be a second nucleus, hidden from our view at
visible and near infrared wavelengths due to significant extinction.
We elaborate on the properties of the second nucleus in section
\ref{secpeak}.  The minor axis dynamics shows a single peak at the
location of the photometric peak, with no other features.
The line of sight recession velocity is remarkably consistent with the
velocities of cold molecular gas observed by \cite{handa}.  
For a galaxy inclination of 24\degr\ to face-on
(\cite{comte}), the deprojected rotation velocities are a factor
of 2.46 higher than the observed values.

We also analyzed the major and minor axis spectra using the Fourier
cross correlation quotient method.  The technique is described by
\cite{bender}, and has been shown to work equally well for asymmetric
line profiles, such as the $^{12}$CO bandhead by \cite{anders},
\cite{tecza} and others.  Essentially, it consists of deconvolving the
galaxy spectrum using an appropriate template star spectrum, so as to
yield the line of sight velocity distribution (LOSVD) for each spatial
resolution element.  The deconvolution is achieved by
cross-correlating the galaxy spectrum with the stellar spectrum, and
dividing the result with the auto-correlation of the stellar spectrum.
The convolutions are carried out in Fourier space, and a Wiener filter
is used to limit the high frequency component of the result, so as to
minimize noise amplification.

The Fourier cross correlation quotient (FCQ) technique is superior to
a simple deconvolution since it suppresses those frequencies in the
result where the template stellar spectrum contains little or no
information.  It has the additional advantage that it is relatively
insensitive to {\em template mismatch}.  The latter occurs when the
spectral type of the template does not match that of the galaxy
spectrum being deconvolved.  The depth of the CO bandhead features is
a function of spectral type and luminosity class (\cite{oliva}), and
$\chi^2$ fitting techniques are prone to interpreting a deeper
template feature as high velocity dispersion in the galaxy spectrum,
and vice versa.  FCQ techniques also make no assumption regarding the
LOSVD being observed.  The disadvantage of the FCQ technique is that
it requires higher SNR spectra, precisely because it makes no apriori
assumptions about the LOSVD.  A cross-check of our dynamical analysis
using FCQ techniques assuming a Gaussian LOSVD yielded almost
identical results to those presented in figure \ref{velprofile}.
Single component, broad LOSVDs were observed at the location of both
dynamical peaks.

\subsection{Stellar population analysis: Extinction and luminous mass
estimates}

\subsubsection{Magnitudes and colors}

We have derived K band magnitudes for the nucleus, several star
forming knots within the near infrared arc, and points along the mid
infrared bar, based on the calibrated NICMOS K band image.  A radial
profile analysis of several of the star forming knots yielded a FWHM
of $\sim$3 pixels (0\farcs 225), consistent with the diffraction limit
at 2.2 \micron .  Consequently, we used an aperture with a radius of
5 pixels for our aperture photometry.  The observed K band magnitudes,
and positions relative to the photometric peak of several knots
are listed in Table \ref{magntable}.  We adopt the nomenclature of
\cite{gallais}, further extended by \cite{elmegreen}, for
consistency.  Note, however, that source 5 (location of SN 1968L,
\cite{wood}) is resolved into two knots,
5a and 5b, by NICMOS, and source 6 (also referred to as source B by
\cite{gallais}) also splits up into two sources 6a and 6b at HST
resolution.  
Source 9 corresponds
to compact emission $\sim$2\arcsec\ north-west of source 5, which is
only detected in the high resolution NICMOS images. 

\begin{table}
\begin{center}
\caption[]{Magnitudes and colors for emission peaks}
\label{magntable}
\begin{tabular}{|c|c|c|c|c|c|}
\hline
\noalign{\smallskip}
Source & $\Delta$RA & $\Delta$Dec & m$_{\rm K}$ &
m$_{\rm H}$ & A$_{\rm V}$ \\
name & from 1 & from 1 & (r=5 pix)&(r=5 pix)& \\
\noalign{\smallskip}
\hline
\noalign{\smallskip}
1  &   0\fs 00 &  0\farcs 00 & 12.31 & 12.63 & 0.9 \\
3  & -0\fs 039 & -7\farcs 03 & 15.27 & 15.56 & 0.5 \\
4  & -0\fs 359 & -4\farcs 76 & 14.44 & 14.86 & 2.5 \\
5a & -0\fs 473 & -4\farcs 00 & 13.40 & 13.83 & 2.6 \\
5b & -0\fs 521 & -3\farcs 55 & 14.22 & 14.73 & 3.8 \\
6a & -0\fs 771 & 4\farcs 57  & 13.86 & 14.35 & 3.5 \\
6b & -0\fs 723 & 4\farcs 48  & 14.60 & 15.30 & 6.7 \\
8  & -0\fs 595 & 1\farcs 86  & 14.59 & 15.45 & 9.2 \\
9  & -0\fs 547 & -2\farcs 40 & 14.03 & 14.44 & 2.3 \\
10 & -0\fs 333 & 0\farcs 50  & 15.32 & 15.67 & 1.4 \\
\noalign{\smallskip}
\hline
\end{tabular}
\vspace{-0.5cm}
\end{center}
\end{table}

The H-K colours listed in table \ref{magntable} have been corrected
for a Galactic extinction corresponding to E(J-K)$=$0.017 mag
(\cite{deVaucouleurs}).  Comparing the observed colours with those
expected for an M type giant/super-giant population (H-K$=$0.25
\cite{korneef}), and assuming a screen model for dust extinction
following the standard Galactic law (\cite{cox}), we find that the
extinction varies over a wide range within the nuclear region.

The nucleus and the southern part of the star forming arc appear only
slightly extincted, with increasing extinction toward the south-west
part of the star forming arc.  Source 6 (10 micron source B) shows a
very strong extinction gradient from 6a to 6b, leading us to postulate
that we are only observing a small part of the star forming activity
which occurs close to the near edge of the cloud.  Source 8 (10 micron
source A) shows the highest extinction, consistent with the conclusion
of \cite{gallais} that it is younger than B.  Most of the region
identified as the mid infrared bar by \cite{gallais} is severely
extincted (A$_{\rm V}$ $\sim$ 10 mag, see also Turner et al. 1987),
including the location of the secondary nucleus.  The high extinction
is hiding most of the star formation activity within the bar from our
view, even at near infrared wavelengths.  The NICMOS Pa $\alpha$ image
shows very strong emission all along the mid infrared bar, with a peak
located near source 10 (table \ref{magntable}), confirming the
presence of a large number of ionizing photons within the region.  The
Pa $\alpha$ emission peak also corresponds to an H $\alpha$ peak
observed in HST WFPC narrow band images (archival data, P.I. S. Heap,
Proposal I.D. 1213).
Ishizuki (1993) observed a large concentration of cold molecular gas
(map published by \cite{sofue}) in the nuclear region of M\,83, with a
peak located in the close vicinity of the mid infrared bar.  

\subsubsection{Equivalent width, age and luminosity}
Amongst all the K band emission peaks identified by \cite{gallais} and
mentioned in table \ref{magntable}, only the nucleus (source 1) lies
within our major and minor axis slits.  We measure an equivalent width
for the $^{12}$CO bandhead of 10.4$\pm$0.4\AA\ at the nuclear location, using
the range 2.2931 -- 2.2983 \micron , as defined by \cite{kleinmann}
(also used by \cite{origlia}). Origlia et al. (1993) derive a correction for
the observed equivalent width as a function of the velocity dispersion
of the source.  We have carried out a similar calibration for the
ISAAC spectra at R $\sim$ 4750.  We artificially broadened template
star spectra using a Gaussian broadening function with widths varying
from $\sigma =$ 0 to 250 km\,s$^{-1}$.  We then measured the CO
equivalent width over the specified bandpass.  A linear least squares
fit to the observed variation of equivalent width with velocity
dispersion yields
\begin{equation}
W_{\rm true} = W_{\rm obs} \times \left( 1 + 1.91 \times 10^{-3}
\sigma \right).
\end{equation}
Our calibration does not show any flattening at low $\sigma$ values,
in contrast to the relationship derived by Origlia et al. (1993).  We
attribute their observed flattening to the lower resolution of their
spectra, since a change in equivalent width would only occur if the
broadening caused by the velocity dispersion is comparable to the
intrinsic instrumental resolution.  We derive a corrected nuclear
equivalent width of 11.9$\pm$0.5\AA , using our formula and the observed
nuclear velocity dispersion (figure \ref{velprofile}). 

We used the population synthesis models by \cite{sternberg} to
estimate the age of the nuclear star formation activity.  We assumed a
single burst of star formation decaying exponentially with a scaling
time to 10$^6$ years, a Salpeter IMF from 1 to 100 M$_{\sun}$, and
solar metallicity for the modeling.  The observed CO equivalent width
would correspond to a burst of star forming activity between 25 and 60
million years ago.  The observed K band luminosity would correspond a
total cluster mass of 2.5 $\times$ 10${^6}$ M$_{\sun}$, assuming a
distance to M\,83 of 3.7 Mpc (\cite{deVaucouleurs79}).  The nuclear K
band light appears to be dominated by a population of giant stars.  We
repeated our analysis using the models of \cite{leitherer}, obtaining
similar results.  The observed equivalent width at the secondary
nucleus, corrected for velocity dispersion, is 11.9$\pm$0.5\AA ,
indicative of a giant population with age similar to the visible
nucleus.  

The star forming arc, in contrast, shows a significantly higher CO
equivalent width of 13.7$\pm$0.5\AA .  The velocity dispersion at this
location 7\arcsec\ from the nucleus along the major axis slit is only
18 km\,s$^{-1}$, yielding a corrected value of 14.2$\pm$0.5\AA .  We
conclude that the K band light from the star forming arc is dominated
by super-giant stars, although we note that the photometric peak along
our major axis slit lies between sources 4 and 5.  The large
equivalent width also implies a younger age of 10 million years for
the starburst in the arc, assuming the same population synthesis model
parameters as for the nucleus.  The observed very low value of
velocity dispersion within the arc is consistent with a picture in
which the star forming activity within the arc is very young, and has
not reached dynamical equilibrium with the gravitational potential of
the galaxy.  The K band light we observe is dominated by the super
giant stellar population which has an internal velocity dispersion too
small to be accurately measured at our instrumental
resolution. \cite{gallais} also confirm that the observed colours of
the star forming arc are consistent with a reddened giant/supergiant
population.

\section{The nature of the second nucleus \label{secpeak}}
The two peaks in the stellar velocity dispersion profile correspond to
dynamically hot systems, if we assume that the stars used to trace the
gravitational potential potential are dynamically relaxed.  It is
unlikely that we are seeing a very young (and potentially unrelaxed)
star cluster, as the observed equivalent widths are those observed for
giant rather than super-giant stars.  In any case, a very young
cluster would exhibit a very low velocity dispersion, in contrast to
the observed peaks.  A second possible mechanism to create such
dynamical peaks without mass concentrations is via velocity anisotropy
(orbit crowding or streaming motions), but such a large effect
(FWHM $\sim$ 175 km\,s$^{-1}$) has not been observed in any galaxy
nucleus.  M\,83 is inclined at 24$\degr$ to the plane of the sky,
making it implausible for any rotation within the galactic plane to
cause the observed peak (required deprojected velocity 430
km\,s$^{-1}$.  We conclude that both peaks likely represent dynamically hot
{\em nuclei}.

If we assume that the stellar system is an isothermal sphere, we
derive an enclosed mass of 1.3 $\times$ 10$^7$\,M$_{\sun}$ within 5.4
pc, using the Jeans equation.  The mass estimate is even higher if we
use the Virial or Bahcall-Tremaine estimators for a system dominated
by a point mass.  The observed mass could be either in stars or a dark
mass concentration.  For the visible nucleus (K band photometric
peak), we derive a mass estimate of 2.5 $\times$ 10$^6$\,M$_{\sun}$
for the stellar component, using population synthesis models.  The
rest of the mass could conceivably exist as a dark mass.  A strong
parallel may be drawn with the Milky Way, where \cite{genzel} have
observed a central dark mass of 2.8 $\times$ 10$^6$\,M$_{\sun}$, and
the enclosed mass at a radius of 5 pc is 1.5 $\times$
10$^7$\,M$_{\sun}$.
 
No such conclusion about the mass concentration at the second nucleus
can be made, as it is substantially extincted and we are unable to
make an accurate estimate of its intrinsic K band luminosity.
However, we note that the location of the second nucleus coincides
almost exactly with the center of symmetry of the outer isophotes of
M\,83, which ought to represent the dynamical center of the galaxy.

Telesco et al. (1993) observe a bar like morphology
of the mid infrared emission.  Elmegreen et al. (1998), using extinction
maps, argue for the presence of a bar within a bar, with the inner bar
orthogonal to the outer one.  Such structures have been predicted and
observed in several galaxies(\cite{maciejewski}, \cite{erwin}).
However, in no case does the inner bar appear to be offset from the
nucleus as it does in M\,83.  Dynamical arguments would place the
galaxy nucleus exactly along the inner bar, as is the case for the
second nucleus observed by us.  The outer dust ring seen by
Elmegreen et al. (1998), possibly associated with the inner Lindblad
resonance, also appears to be centered on the location of the second
nucleus.  The visible nucleus might, together with the star forming
arc and source B, form a ring of star formation activity centered at
the location of the second nucleus.

The nearest normal spiral galaxy, M\,31, also shows evidence for a
double nucleus (\cite{kormendy}; \cite{statler}; \cite{bacon}).  While the
nature and cause of the double nucleus is not well understood,
transient phenomenon, such as interaction with a companion, could be
responsible (\cite{bacon}).  Interaction between NGC 5253 and M\,83
has been postulated by \cite{wolstencroft}, it could also be
responsible for triggering the starburst activity in M\,83.  The
nuclear region is also a strong X ray source (\cite{immler}), although
the spatial resolution available to date is inadequate to identify
whether it is associated with either of the two dynamical peaks.  

\section{Conclusions}
We have carried out medium resolution long slit spectroscopy along the
major and minor axis of M83, covering the nuclear region.  Using the
deep, sharp CO bandhead features longward of 2.29\micron , we have
analyzed the dynamics of the stellar population in the nuclear region,
measuring the recession velocity and velocity dispersion.  The
velocity dispersion profile shows two peaks, one located at the
photometric peak of K band light, and the other offset 2\farcs 7 south
west of it.  Each of the two peaks imply an enclosed dynamical mass of 1.3
$\times$ 10$^7$\,M$_{\sun}$ within 5.4 pc, if the stellar population
is dynamically relaxed.  The K band emission observed toward the
photometric peak is dominated by light from giant stars, with ages
between 25 and 60 million years, estimated using population synthesis
models for an instantaneous burst.  The estimated total mass of such a
stellar cluster is 2.5 $\times$ 10$^6$ M$_{\sun}$.  

The off-nuclear dynamical peak might correspond to a mass
concentration located at the dynamical center of M\,83, as evidenced
by the center of symmetry of stellar isophotes.  Very little K band
emission is observed at the location of the second nucleus.  It is
likely that it is hidden from our view by $>$10 magnitudes of
extinction, since its position lies within a bar of mid-infrared
emission and high extinction.  We postulate that the bar represents
gas flowing toward the dynamical center of the galaxy.  Star formation
triggered within the bar is likely responsible for the observed
mid-infrared emission.

\begin{acknowledgements}
We would like to thank the ESO Paranal staff, especially Jean-Gabriel Cuby
and Gianni Marconi, for extensive help with the ISAAC
observations. The ESO La Silla and Garching staff deserve credit for
carrying out the SOFI observing in service mode.  We thank the referee
for insightful comments.  This research made
use of the STSCI HST archive, the ESO ST-ECF archive, and the NED
and SIMBAD databases.  We thank them all for the support. 
\end{acknowledgements}

\end{document}